\documentclass[preprint,prb]{revtex4}
\usepackage{amssymb}
\usepackage{amsmath}
\usepackage{epsfig}

\def\bra#1{\mathinner{\langle{#1}|}}
\def\ket#1{\mathinner{|{#1}\rangle}}

\def\bracket#1#2#3{\langle#1\vert#2\vert#3\rangle}
\def\expect#1{\langle#1\rangle}
\def\leer{\varnothing}

\def\id3{ {\mathbf{1}_{3\times 3}}}
\def\ln {\text{ln}}

\begin{document}

\title{Dependence of the viscosity on the chain end dynamics in polymer
melts}
\author{Matthias Pae\ss ens}
\email{m.paessens@fz-juelich.de}

\affiliation{Institut f\"ur Festk\"orperforschung, Forschungszentrum J\"ulich - 
52425 J\"ulich, Germany}
\date{\today}

\begin{abstract}
We compare the Rubinstein--Duke model for reptation to a model where the boundary dynamics is modified
by calculating the viscosity of polymer melts. The question is investigated whether the viscosity
is determined by details of the dynamics of the polymer ends or by the 
stretching of the polymer. To this end
the dependence of the viscosity on the particle density of the lattice gas models which can
be identified by the stretching is determined. We show that the influence of the stretching of the
polymer on the absolute value of the viscosity in the scaling limit of of very long chains 
is much bigger than the influence of the boundary
dynamics, whereas the corrections of the scaling of the viscosity depends significantly 
on the details of the boundary
dynamics.

\end{abstract}

\maketitle

\section{Introduction}
A persistent problem in polymer science is the theoretical description 
of the mass dependence of the viscosity
of polymer melts of linear chains.\cite{mcleish} A model which recently lead to some
progress\cite{carlon,carlon2,ps} is the
Rubinstein--Duke (RD) model,\cite{rubinstein,duke} which is a discrete model that maps the reptation
dynamics of a three dimensional polymer onto a lattice--gas.
This model takes into account tube length fluctuations as well
as the effect of a possible external electric field which acts on a charged polymer.\\
In the reptation model the stretching of the polymer is caused by an entropic tensile force
acting on the chain ends.\cite{doiedwards} In the Rubinstein model this force is parameterized
by the lattice dimension which provides the rates for adding and removing new segments
to the tube. Adding (resp. removing) new segments is modeled in the lattice--gas picture by 
the creation (resp. annihilation) of particles at the boundaries.
The stretching
of the polymer corresponds to the particle density in the bulk. Although the 
real spatial dimension of the polymer network 
(naturally $d=3$) determines the possibilities of the polymer to stretch and by this it is
directly related to the creation rate, this rate is not only influenced by the dimension because the
dynamics of the boundary segments of a polymer is affected by more mechanisms than pure
geometry. \\
Rubinstein investigated the influence of the creation and annihilation rates on the viscosity which
is proportional to the relaxation time and showed that the magnitude of the viscosity 
increases with increasing creation rates. 
Due to particle number conservation in the bulk the particle density is provided 
by the boundary dynamics -- none the less the density is a property of the bulk and 
different boundary dynamics may lead to the same density. This feature of the model
corresponds to a realistic property of the polymer: If the behavior of the boundary segments
is changed, the stretching of the polymer will not necessarily be affected. 
In order to investigate whether the
boundary dynamics or the particle density determines the relaxation time we consider two
models: The standard Rubinstein--Duke model and a modified
version. The model is only modified at the boundaries in a way that in the bulk
the particle density and thus the entropic force remains unchanged.\cite{euroschuetz}\\
We will show that the particle density is the crucial property, so that it is reasonable to 
specify the density dependence of the relaxation time explicitly. As a hypothesis we convert
the known density dependence of the diffusion constant to an expression for the relaxation time.
This hypothesis is confirmed by our numerical data implying that 
we have on the one hand determined the density
dependence and on the other hand verified the conversion of the diffusion constant to the
relaxation time. However, we also show that microscopic details of chain end dynamics have an 
appreciable impact on the absolute value of the viscosity and its scaling.

\section{Model}
\subsection{Standard Rubinstein--Duke model}
\begin{figure}
{   \begin{center}
\resizebox*{!}{3.8cm}{\includegraphics{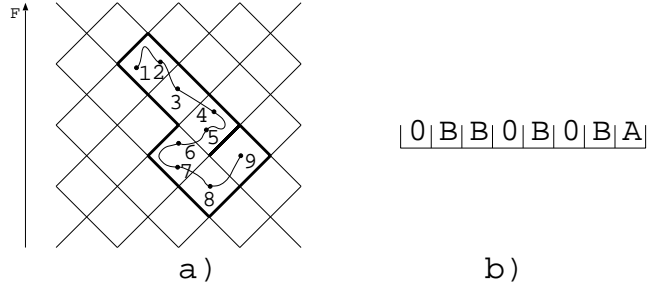}}
    \end{center} }
\caption{a) The repton model in two dimensions: the circles represent the reptons; the
primitive path is marked by the bold lines. b) Projection onto one dimension.}
\label{reptonmodel}
\end{figure}

Rubinstein assumes that the constraints of the other polymers divide space into cells
which form a $d$--dimensional regular cubic lattice. The polymer occupies a series of
adjacent cells, the ``primitive path''. It is not possible for the polymer to traverse
the edges of the cells (in two dimensions: the lattice points) so that only the ends
of the polymer can enter new cells. The polymer is divided into segments whose lengths
are of the order of the lattice constant, the number of segments is proportional
to the length of the polymer or the molecular weight. The ends of the segments are labeled
with imaginary particles called ``reptons''. The orientation of the lattice
is introduced by Duke in a way that the electric field is diagonal to the lattice, i.e.
in three dimensions the (111)--direction, which is relevant e.g. for the investigation
of electrophoresis. In this paper we will only consider the limit 
of zero field -- nevertheless we
will keep a reference axis along which the displacement
of the polymer chain as a whole can be monitored.\cite{barkema1,barkema2}\\
A repton is allowed to jump into an adjacent cell according to the 
following rules: \\
\begin{enumerate}
\item The reptons in the bulk are only allowed to jump along the primitive path.
\item No cell in the interior of the primitive path may be left empty.
\item The ends move freely provided rule 2 is respected. If an end repton occupies 
the cell alone, it can only retract in the cell of the adjacent repton. If the 
adjacent repton is in the same cell, the end repton may enter any of the $2d$ 
surrounding cells.
Reptons in the bulk jump with the same probability as 
reptons at the end into occupied cells.
\end{enumerate}
Rule 1 refers to the diffusion of defects and ensures that the polymer does not traverse 
the topological constraints which are
represented by the edges of the cells in the model. Rule 2 refers to the connectivity of
the polymer and 
is motivated by the fact that the segments are of the size of the lattice constant and are not
allowed to stretch. 
Finally rule 3 reflects the fact that there are more free adjacent cells for an 
end repton than occupied ones -- this introduces the entropic tensile force. \\
As the shape of the primitive path is not affected by the movement of the polymer 
and only the ends are created or annihilated this model can be mapped onto one 
dimension. To this end a particle representation is used: 
Starting on one end, a segment into the direction of the reference axis is identified with 
an $A$--particle, against the axis with a $B$--particle and finally a segment within one cell by a vacancy 
$\varnothing$ (Fig. \ref{reptonmodel}).
\\
These particles are residing on a chain whose number of sites $L$ is the number of 
segments, thus the number of reptons minus one. As for each site three states $\{A$,$B$,$\varnothing \}$ are 
possible this model can be identified with a quantum spin--one chain.
\cite{schuetz,barkema0}\\
The one--dimensional dynamics of the particles is the following: In the bulk, 
$A$-- and $B$--particles interchange freely with vacancies $\leer$, but may 
not traverse each other. By these rules it is guaranteed that the shape of the primitive
path is conserved. \\
At the boundaries, particles may be annihilated with rate $\beta$, i.e. transformed into vacancies,
or created with rate $\alpha$, i.e. the transformation of vacancies into particles. Within the approach of
Rubinstein the annihilation
is a process with the same rate as the hopping in the bulk which is a basic process
(rate is assigned to $1$), i.e. $\beta=1$. The creation of 
particles is $d$--times more probable, because at the boundaries
there are $2d$ cells to enter, $d$ along and $d$ against the direction of the field, while
in the bulk there is only one cell to enter. Hence according to Rubinstein, $\alpha=d$.\\
In this way a stochastic interacting particle system on a one dimensional chain has 
been defined. As the transitions are independent of the previous history this process
is Markovian.
A convenient way to describe the process mathematically is the quantum Hamiltonian 
formalism which we will present here briefly, for details see Ref. 
\onlinecite{schuetz}.\\
Assigning to the particles and the vacancy one of the three-dimensional basis
vectors each, a configuration of the spin chain can be represented by the tensor product
$\ket\eta$ of these vectors. The probability to be in state $\ket\eta$ at time $t$ is
labeled by $P_\eta(t)$. These probabilities of the individual states can be combined
to a vector: $\ket{P(t)}=\sum P_\eta(t)\ket\eta$. Due to the conservation of probability
the entries of the vector $\ket{P(t)}$ sum up to 1 at any time $t$. 
With this definition the master equation can be written as: 
\begin{equation*}
\frac{d}{dt}\ket{P(t)}=-H\ket{P(t)},
\end{equation*}
with the stochastic generator 
\begin{equation*}
H= -\sum_\eta\sum_{\eta^\prime\neq\eta} w_{\eta^\prime \rightarrow \eta} \ket\eta\bra{\eta^\prime}
    + \sum_\eta\sum_{\eta^\prime\neq\eta}w_{\eta \rightarrow \eta^\prime} \ket\eta\bra{\eta}.
\end{equation*}
Here $w_{\eta^\prime \rightarrow \eta}$ is the transition probability from state 
$\ket{\eta^\prime}$ to $\ket{\eta}$. In other words the off diagonal elements of 
the matrix $H$ are the negative transition rates between the respective states and 
the diagonal elements are the sum of the rates leading away from the respective state.\\
The creation operators $a^\dagger$ and $b^\dagger$ are defined by $a^\dagger \ket{\leer} =\ket{A}$ 
and $b^\dagger \ket{\leer} =\ket{B}$, acting on any other state yields zero. 
The annihilation operators $a$ and $b$ are defined by $a\ket{A}=\ket{\leer}$ and $b\ket{B}=\ket{\leer}$, again
acting on any other state yields zero. Finally we define the number operators 
$n^A=a^\dagger a$, $n^B=b^\dagger b$ and $n^\leer=1-n^A-n^B$. By these definitions the 
stochastic generator $H$ of the RD model reads: \cite{barkema0}
\begin{equation}
H=b_1(d,1)+b_L(d,1)+\sum_{n=1}^{L-1}u_n
\label{H_rubinstein_duke}
\end{equation}
with
\begin{eqnarray*}
b_n(\alpha,\beta)&=&\alpha\left[n_n^\leer -a_n^\dagger+n_n^\leer -b_n^\dagger\right]\\
&+&\beta\left[n_n^A-a_n+n_n^B-b_n\right] \\
u_n&=&n_n^An_{n+1}^\leer -a_na_{n+1}^\dagger  +   n_n^Bn_{n+1}^\leer -b_nb_{n+1}^\dagger \\
  &+& n_n^\leer n_{n+1}^A-a^\dagger_na_{n+1}  +   n_n^\leer n_{n+1}^B-b^\dagger_nb_{n+1}.
\end{eqnarray*} \\
As mentioned in the introduction we are interested in the influence of the stretching
of the polymer on the viscosity of the polymer melt. To this end we 
present how to determine the stretching and then how to calculate the viscosity.\\
Each particle contributes a length of the order of the lattice constant $c$ to the
primitive path, so that the length of the primitive path equals $Nc$, where $N$
is the total number of particles on the chain. The stretching of the polymer corresponds
to the length to the primitive path divided by the number of bonds $L$ -- or in the
particle picture to the particle density $\rho$. The particle density $\rho$ is 
determined by the creation rate $\alpha$ and the annihilation rate $\beta$. 
In the bulk the density on {\em one} site changes only due to the hopping process:
 \begin{equation}
\begin{split}
\frac{d}{dt}\expect{n_k} &= \expect{(1-n_k) (n_{k+1}+n_{k-1})} \\
                         & - \expect{(n_k) (1-n_{k+1}+1-n_{k-1})} \\
     &= \expect{n_{k+1}} + \expect{n_{k-1}} - 2 \expect{n_k}.
\end{split}
\label{eq_nk}
\end{equation}
Here, $k=2,\ldots, L-1$ and $n_k=n_k^A+n_k^B$. The first term on the right hand side
equals the gain of particle which is proportional to the occupation of the site 
with vacancies and proportional to the occupation of the adjacent
sites with particles. Accordingly, the second term describes the loss of particles.
Supplementary at the boundaries $k=\{1,L\}$ we have to take creation and annihilation into account:
\begin{equation}
\begin{split}
\frac{d}{dt}\expect{n_k} &= \expect{({\bf 1}-n_k) (n_{k\pm 1})} -
                             \expect{(n_k) ({\bf 1}-n_{k\pm 1})} \\
                          &  + 2\alpha \expect{{\bf 1}-n_k} - \beta \expect{n_k} \\
 &= \expect{n_{k\pm 1}}-\expect{n_k} + 2\alpha - (2\alpha+\beta) \expect{n_k}.
\end{split}
\label{eq_nrand}
\end{equation}
Here, $k\in\{1,L\}$ and the index $k+1$ for $k=1$ respectively $k-1$ for $k=L$.
We have to consider the rate $2\alpha$ because both $A$-- and $B$--particles
are counted by $n_k$. As we are only interested in the stationary state we presume
$d/dt\,\expect{n^\ast_k}=0$ for $k=1,\ldots,L$. Eq. (\ref{eq_nk}) then yields
\begin{equation*}
\expect{n_k^\ast}=a+bk,
\end{equation*}
with constants $a$ and $b$. Finally, Eq. (\ref{eq_nrand}) determines 
$a=2\alpha/(2\alpha+\beta)$ and $b=0$. \\
So the stationary total particle number $N^\ast=\sum_{k=1}^L\expect{n_k^\ast}$
is calculated to\cite{spohn}
\begin{equation*}
N^\ast=\frac{2\alpha}{2\alpha+\beta}L,
\end{equation*}
and for the particle density $\rho^\ast=N^\ast/L$ we get
\begin{equation}
\rho^\ast=\frac{2\alpha}{2\alpha+\beta}.
\label{eq.dichte}
\end{equation}
In the RD model one therefore has $\rho^\ast=\frac{2d}{2d+1}$.
The viscosity is proportional to the longest relaxation time of the
stochastic generator\cite{doiedwards} $H$ which is the inverse energy gap of the spin chain.
A very efficient algorithm to calculate the lowest excitations of quantum 
spin chains is the density matrix
renormalization group (DMRG) algorithm.\cite{white,noack} For the standard RD model
the usefulness of this algorithm has been demonstrated,\cite{carlon,carlon2,ps} so that
we employ this algorithm as well for the modifications of the RD model.
The DMRG algorithm is a numerical method to diagonalize huge matrices whose systematic
is known. This is done by reducing the Hilbert space to the most important states which
are chosen by calculating their probability using the density matrix.

\subsection{Reservoir at the boundaries}
In order to investigate the influence of the boundary dynamics the following modification
of the model is considered: In the bulk the dynamics of the particles remains unchanged.
At the boundaries the creation and annihilation rates are chosen as if the chain was
in contact with a particle reservoir, as described in Ref. \onlinecite{euroschuetz}. In the
reservoir the density of $A$--particles is labeled with $\rho/2$, the density of 
$B$--particles is the same, so that the total density is $\rho$. The creation
of particles at the boundary is identified with the hopping of a particle from a 
boundary site of the
reservoir onto a boundary site on the chain. As the probability to find an 
$A$-- or $B$--particle on a site of the reservoir is $\rho/2$ each, the rate for the
creation process has to be chosen as $\rho/2$. Accordingly, the annihilation of
a particle corresponds to the hopping of a particle from the boundary of the chain
into the reservoir. This hopping is only possible when the boundary site of the
reservoir is empty which happens with the probability $1-\rho$. Therefore, this
is the annihilation rate. Using Eq. (\ref{eq.dichte}) it is easily calculated that
this boundary dynamics provides a particle density $\rho$ in the bulk.\\
If we want to provide the same particle density as in the RD model with lattice dimension
$d$, we therefore have to choose the creation rate $\alpha^\prime=d/(2d+1)$ and annihilation
rate $\beta^\prime=1/(2d+1)$. The modified rates differ from the previous ones by a
factor $1/(2d+1)$. A physical interpretation of this modification of the model is that
the dynamics of the boundary reptons is slowed down, as the rates for creation and 
annihilation are reduced. Experimentally this could be seen for polymers with bulky
end groups.

\section{Calculations and results}
\subsection{Comparison of boundary dynamics}
\begin{figure}{
\begin{center}
\epsfig{figure=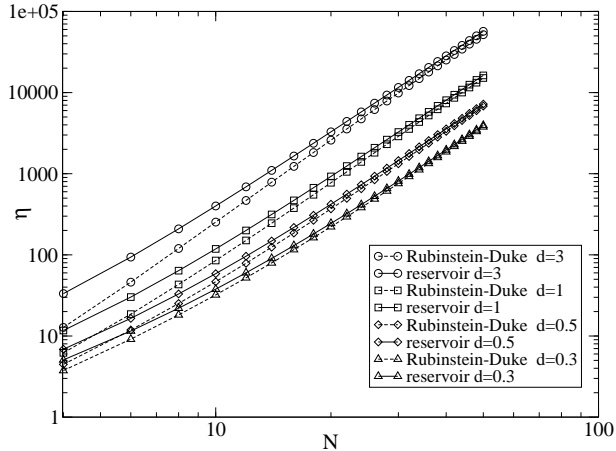,scale=0.3,angle=-90} 
\caption{Viscosity in dependence of the chain length -- comparison of the different boundary dynamics.}\label{eta}
\end{center}
 }
\end{figure}

We calculate in the standard RD model the relaxation times for  $d=3$, $d=1$, $d=0.5$, 
and $d=0.3$. Small values of $d$ seem to be rather artificial -- however, as the
entanglement network is not regular cubic but instead contains also very small
pores, one may investigate smaller effective dimensions than $d=3$.
An illustrative argument is that it is much easier to pull a thread out of the
eye of a needle than to pass it through --- by this the retracting of a polymer should be
more probable than entering a new cell.  \\
For $d=3$ one has a density $\rho=6/7$ so that the length of the
primitive path is $l_{d=3}=0.86cL$ with the lattice constant $c$ and number of bonds
$L$ -- so the polymer is nearly stretched. For the other dimensions the length of the
primitive path calculates to $l_{d=1}=0.67cL$ ($\rho=2/3$), $l_{d=0.5}=0.5cL$ ($\rho=1/2$), 
and $l_{d=0.3}=0.38cL$ ($\rho=3/8$), the length decreases clearly. \\
To compare the standard RD model with the modified boundary dynamics, we
compute the relaxation times in the modified model with the same particle densities. 
The results for both models are plotted in Fig. \ref{eta}.
 \\
With decreasing $d$ the relaxation time decreases in both models. This can be explained
by the fact that a shorter tube relaxes faster. In the particle picture this is justified by
considering that the relaxation process is the traversion of a particle from one end of 
the chain to the other. As for a lower particle density the diffusion is faster this
process takes less time.\\
One can observe two qualitative results: Firstly compared to the RD model 
the relaxation times of the reservoir boundary
dynamics are shifted to higher values. Secondly  it is obvious that the influence 
of the particle density is much stronger than the influence of the boundary dynamics.
\subsection{Density dependence of the relaxation time}
After having verified in the last section that the particle density is crucial for the relaxation
time, in this section the dependence is quantified. To this end a method for computing
the diffusion constant from the relaxation time is presented. As the density dependence of the 
diffusion constant is known\cite{spohn2,newman} this leads to a hypothesis for the density
dependence of the relaxation time. By verifying this hypothesis not only the density dependence
 is found but also the validity of the computation of the diffusion constant is proved.\\
As the RD model is a projection to a specific axis, the diffusion constant along this axis will be
calculated which differs only by a constant factor from the three dimensional diffusion constant.
The projected end to end length of the polymer is the difference of the number of $A$-- and 
$B$--particles, $\left| N_A - N_B \right|$. The one dimensional diffusion can be identified 
with a random walk: After the relaxation time $\tau$ the polymer has moved by a distance 
$\left| N_A - N_B \right|$ in or against the direction of the field, i.e.
\begin{equation*}
x(t+\tau)=x(t)+a_n,
\end{equation*}
with $a_n=\pm \left| N_A - N_B \right|$. Assuming that the random walker starts at $x(0)=0$
the following relation holds (for sufficiently large times $t$ so that the number of steps
is well described by $K=t/\tau$):
 \begin{equation*}
x(t)=\sum_{n=1}^{K=t/\tau}a_n,
\end{equation*}
which yields
\begin{equation*}
\expect{x(t)^2}=\expect{\sum_{n=1}^{K=t/\tau}a_n^2}=t\frac{\expect{\left(N_A-N_B\right)^2}}{\tau},
\end{equation*}
because the $a_n$ are uncorrelated: $\expect{a_n a_m}=\delta_{n,m} a_n^2$. Thus calculating the diffusion
constant yields
\begin{equation*}
D=\frac{\expect{\left(N_A-N_B\right)^2}}{2\tau}.
\end{equation*}\\
As described in Ref. \onlinecite{schuetz} the expectation value is calculated by
\begin{equation*}
\expect{\left(N_A-N_B\right)^2}=\bracket{s}{\left(N_A-N_B\right)^2}{P^\ast},
\end{equation*}
with the ``summation vector'' $\bra{s}=(1,1,...)$ and the stationary state $\ket{P^\ast}$. 
With a product ansatz the stationary state can be computed to\cite{richard} (here with
generalized creation and annihilation rates $\alpha$ respectively $\beta$)
\begin{equation*}
\ket{P^\ast}=\left[  \frac{\beta}{2\alpha+\beta}
              \begin{pmatrix} \frac{\alpha}{\beta} \\ 1 \\ \frac{\alpha}{\beta}\end{pmatrix}
              \right]^{\otimes L},  
\end{equation*}
so that
\begin{equation*}
\expect{\left(N_A-N_B\right)^2}=
  L\frac{2\alpha}{2\alpha+\beta}=L\rho.
\end{equation*}
Thus we get for the diffusion constant the expression
\begin{equation}
D=\frac{L\rho}{2\tau}.
\label{d_von_tau}\end{equation}\\
In Ref. \onlinecite{spohn2} or \onlinecite{newman} it is shown that the density dependence of
the diffusion constant is
  \begin{equation}
D=\frac{1}{L^2}\frac{1}{2d+1}K(L)=\frac{1}{L^2}\left(1-\rho\right)K(L).
\label{D_rho}\end{equation}
The term $K(L)$ corrects finite size effects of the length dependence. Thus
the density dependence of the relaxation time should be
\begin{equation}
\tau=L^3\frac{\rho}{1-\rho}\frac{1}{2K(L)},
\end{equation}
so that the expression
\begin{equation}
\frac{L^3}{\tau}\frac{\rho}{1-\rho}=2K(L)
\label{eqN3E}
\end{equation}
should be independent of $\rho$. In order to avoid length corrections the limiting case $L\to\infty$
is considered by extrapolating the term $L^3/\tau$ by the algorithm of Bulirsch and Stoer (BST)
as described in Ref. \onlinecite{BST}.  
\begin{figure}{
\begin{center}
\epsfig{figure=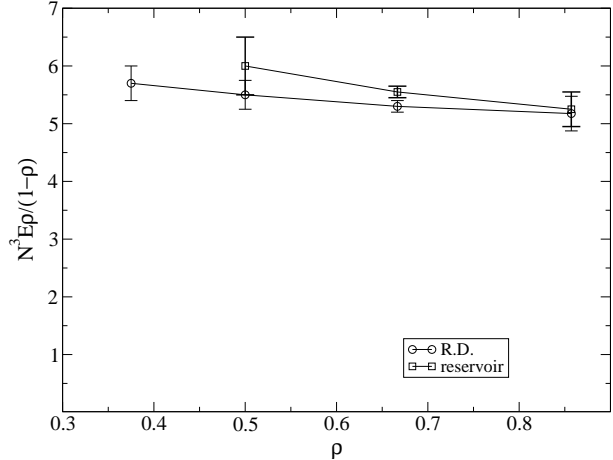,scale=0.3,angle=-90} 
\caption{The extrapolated value $\lim_{L\to\infty}L^3E\rho/(1-\rho)$ vs. $\rho$. For both boundary
dynamics the result is consistent with the predicted density dependence.}\label{n3erho}
\end{center}
 }
\end{figure} \\
In the BST algorithm a degree of freedom $\omega$ has to be chosen whose size is of the order
of the correction exponent of the extrapolated series, but whose influence on the result of
the extrapolation is usually not critical. For the plotted data $\omega=1.5$ was used, because
in this region a stable area could be found. Only for the reservoir with $\rho=0.375$ no
convergence of the algorithm could be achieved. The error of the extrapolation is determined
by a variation of $\omega$ in the stable area. \\
Contrary to expectations the extrapolated values show some dependence on the parameter $\omega$.
For $\omega=0.5$ which is the assumed correction exponent\cite{doiedwards} lower values are found 
but a significant $\rho$ dependence can not be found.\\
The results for the different boundary dynamics and particle
densities are plotted in Fig. \ref{n3erho}. In consideration of the error bars no density dependence
can be found so that we have found the correct density dependence. 
The density dependence found here confirms the dependence 
conjectured by Rubinstein,\cite{rubinstein}
who proposes a density dependence of $\tau\propto z(z-1) N^3$, where $z-1$ is the number of
cells which can be entered by an end repton (which is $2d$ in our notation). The factor $z-1$ corresponds
exactly to $\rho/(1-\rho)$ and the additional factor $z$ results from the fact that our time
scale is $z$--times the time scale used by Rubinstein, so that every time measured in our scale
must be multiplied by a factor of $z$ when being compared to the Rubinstein scale.     
We have thus shown that the conjecture is not only valid for the Rubinstein model but also 
for generalizations.

\subsection{Effective Exponent, details of chain end dynamics}
\begin{figure}{
\begin{center}
\epsfig{figure=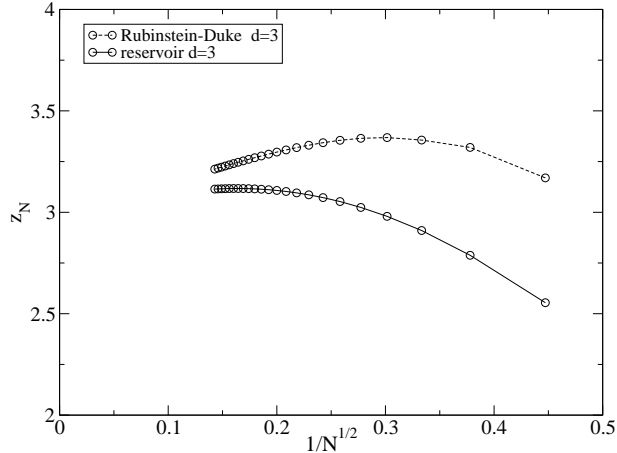,scale=0.3,angle=-90} 
\caption{Length dependence of the effective exponent -- influence of the boundary dynamics, $d=3$ i.e. $\rho=6/7$.}\label{exp}
\end{center}
 }
\end{figure}

As in polymer physics the scaling of the viscosity with the polymer weight $M$ is 
investigated intensively we will now focus directly on this quantity. Experiments
show a scaling of $\eta\propto M^{3.4}$ for several magnitudes of polymer weight and
a crossover to an exponent of $3$ in the limit of infinitely long chains.\\
The effective exponent of the scaling is given by the local slope $z_N$ 
in the log--log--plot,
which is calculated by
\begin{equation}
z_N=\frac{\ln \tau_{N+1} - \ln {\tau_{N-1}} }{ \ln (N+1) - \ln (N-1)}.
\end{equation} 
Fig. \ref{exp} shows the effective exponent for both boundary dynamics with $d=3$. As
abscissa values we have chosen $N^{-1/2}$ because according to Doi the 
finite size correction should
be of this order. Compared to the RD model the reservoir--dynamics shifts the effective
exponent to smaller values, with increasing chain length the deviation decreases. While
the RD model shows an exponent in the region which is found experimentally at least 
in a small range, the values found for the reservoir--dynamics are too small. But using
once again the BST--algorithm it can be shown that both curves approach $z_N=3$ for 
$N\to\infty$: for $d=3$ we get for the RD model $\lim_{N\to\infty} z_N=3.03\pm 0.07$ 
and for the reservoir--dynamics  $\lim_{N\to\infty} z_N=3.00\pm 0.03$. For other values
of $d$ the same qualitative results can be found, but for smaller values of $d$ the
difference of the curves decreases.\\
So we conclude that although the influence of the boundary is much smaller than the
influence of the parameter $d$, for some quantities the type of boundary dynamics
is crucial. We
see that the effective exponent is a very sensitive quantity for which the type of
model which is used for the chain end dynamics is of big influence. 

\section{Summary}
In this paper we have investigated the influence of the boundary dynamics on the viscosity of
polymer melts by comparing the Rubinstein--Duke model to a model with modified boundary
dynamics. The model is modified in a way that the entropic force which leads to the 
stretching of the polymer remains the same.
In the lattice gas picture this is achieved by introducing particle reservoirs at the boundaries
which provide the required particle density corresponding to 
the stretching of the polymer.\\
We find that the influence of particle density and hence the influence of the entropic force 
is bigger than the influence
of the details of the chain end dynamics. The dependence of the viscosity on the particle
density is quantified in general recovering the conjecture of Rubinstein as a special case. 
Focusing on the scaling of the viscosity we observe that the exponent seen in
experiments is recovered much better with the Rubinstein--Duke dynamics than with 
its modifications. Nevertheless both models lead to the same exponent of 3 in the
limit of infinitely long chains. So although the chain end dynamics has smaller influence 
on the viscosity of polymer melts than the particle density it nevertheless affects considerably
the scaling of the viscosity.

\section{Acknowledgments}
G.M.~Sch\"utz is gratefully acknowledged for the introduction into the topic and many helpful
discussions. The Deutsche Forschungsgemeinschaft is acknowledged for financial support.

\end{document}